\documentclass[10pt,twocolumn,letterpaper]{article}
\pdfoutput=1
\usepackage{times}
\usepackage{epsfig}
\usepackage{graphicx}
\usepackage{amsmath}
\usepackage{amssymb}
\usepackage{algorithm}
\usepackage{algorithmic}
\usepackage{multirow}
\usepackage{subfigure}
\usepackage[pagebackref=true,breaklinks=true,letterpaper=true,colorlinks,bookmarks=false]{hyperref}

\begin{document}

\title{A Low-rank Tensor Dictionary Learning Method for Multi-spectral Images Denoising}

\author{Xiao Gong and Wei Chen\\
Beijing Jiaotong University\\
\small
Code is available at https://www.dropbox.com/s/80ltjrxr2v9maeg/LTDL.zip?dl=0\\
}

\maketitle

\begin{abstract}
As a 3-order tensor, a multi-spectral image (MSI) has dozens of spectral bands, which can deliver more information for real scenes. However, real MSIs are often corrupted by noises in the sensing process, which will further deteriorate the performance of higher-level classification and recognition tasks. In this paper, we propose a Low-rank Tensor Dictionary Learning (LTDL) method for MSI denoising. Firstly, we extract blocks from the MSI and cluster them into groups. Then instead of using the exactly low-rank model, we consider a nearly low-rank approximation, which is closer to the latent low-rank structure of the clean groups of real MSIs. In addition, we propose to learn an spatial dictionary and an spectral dictionary, which contain the spatial features and spectral features respectively of the whole MSI and are shared among different groups. Hence the LTDL method utilizes both the latent low-rank prior of each group and the correlation of different groups via the shared dictionaries. Experiments on synthetic data validate the effectiveness of dictionary learning by the LTDL. Experiments on real MSIs demonstrate the superior denoising performance of the proposed method in comparison to state-of-the-art methods.
\end{abstract}

\section{Introduction}
\label{sec:into}
A multi-spectral image (MSI) has dozens of spectral bands, where the wavelengths may range from infrared to ultra-violet. Compared with a RGB image which only has three spectral bands, an MSI provides more information which reveals features of the object hidden in the spectral domain. However, in many cases, MSIs suffer from corruptions or noises in the sensing process~\cite{aiazzi2001information}. As a low level image processing technique, MSI denoising is key to many high-level computer vision tasks, such as segmentation and classification whose performance highly relies on the quality of the data.

\par
As a model driven approach, dictionary learning methods have been used to find the basic atoms which comprise various signals of a training dataset.
By using a learned dictionary of some signal ensembles, noises can be effectively removed via solving a sparse signal recovery problem for each patch of an image~\cite{elad2006image,dong2011sparsity}. For MSI denoising, applying the traditional dictionary learning methods, e.g., K-SVD~\cite{aharon2006k}, for each band leads to the poor performance, as it fails to exploit spectral information in MSIs~\cite{peng2014decomposable}.

\par
Tensor dictionary learning, which keeps the multidimensional structure of tensors, has attracted growing interests of researchers to process images in the past years. Based on CANDECOMP/PARAFAC (CP) decomposition, Duan et al. extend the K-SVD method for tensors, where a higher order tensor dictionary is learned and each atom of the dictionary is a rank-one tensor~\cite{duan2012k}. By using the Tucker model of tensors, Zubair and Wang propose to learn multiple orthogonal dictionaries along different modes of tensors, where the core tensor have sparse non-zeros elements~\cite{zubair2013tensor}. In~\cite{qi2016tensr}, Qi et al. divide an MSI into small 3-order tensor blocks, and learn overcomplete dictionaries for each mode of the blocks via a two-phase block-coordinate-relaxation approach that includes sparse coding and dictionary updating. However, they fails to further employ all the structural information embedded in images.

In~\cite{dabov2007image,maggioni2013nonlocal}, non-local similar small patches in space are clustered into groups and processed together, which improves the image denoising performance. In addition, in view of the fact that blocks extracted from MSIs have both the spatial correlation and the spectral correlation, the low-rank model is employed in different dictionary learning methods for MSI denoising. For example, Peng et al. apply dictionary learning for each group (i.e., a tensor) separately and enforce a low-rank Tucker approximation~\cite{peng2014decomposable}. A simultaneously sparse and low-rank structure is considered for each group of an MSI in~\cite{xie2017kronecker}. In both two methods, each group of an MSI is processed separately and a learned dictionary only captures information of the group, which fails to exploit inter-group correlations. Learning a shared dictionary among all groups of an MSI would gain from the multitask learning concept and be more effective to capture the atoms that comprises the signal.

\begin{figure*}[t!]
\begin{center}
\includegraphics[width=0.9\linewidth]{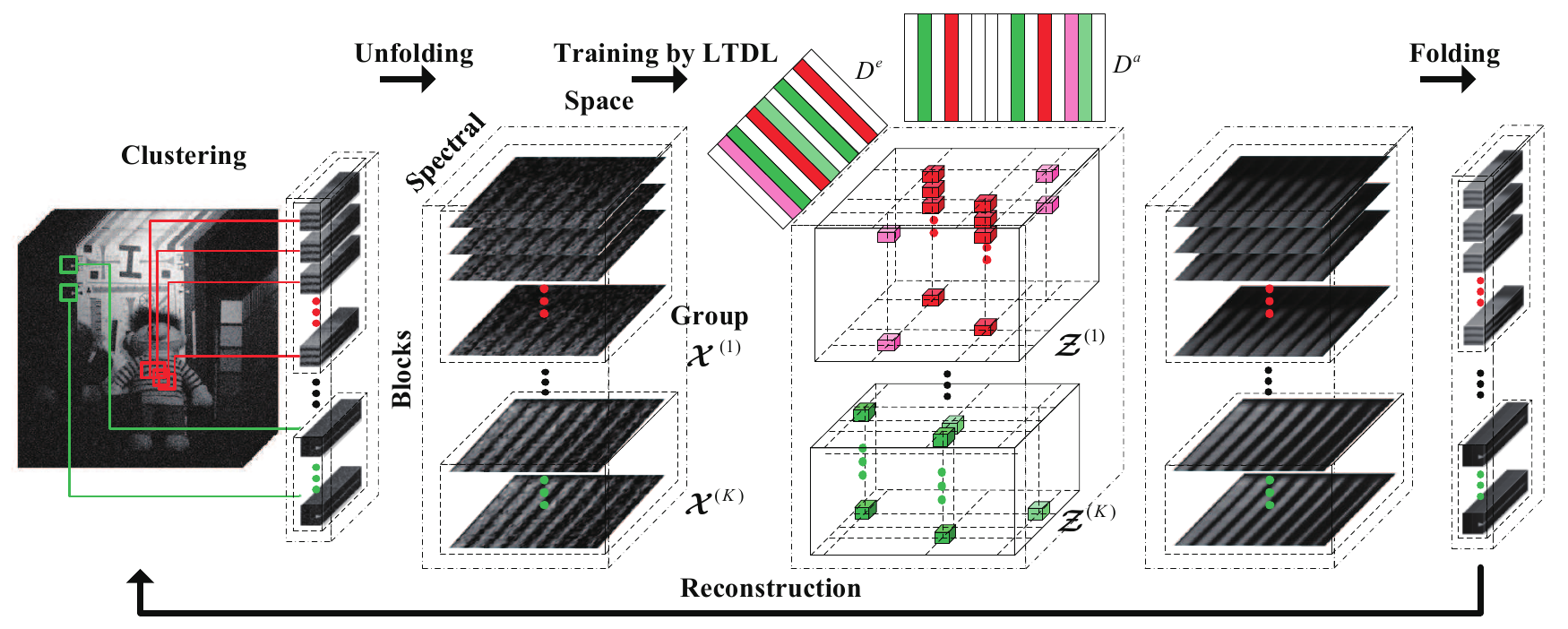}
\end{center}
   \caption{Flowchart of the proposed MSI denoising method.}
\label{fig:fig1}
\end{figure*}

In this paper, we propose a Low-rank Tensor Dictionary Learning (LTDL) method for MSI denoising, which differs with existing methods in two aspects:
\begin{itemize}
\item Dictionaries of the spatial domain and the spectral domain are trained by all tensor groups in the proposed method, so the features in space and spectrum can be learned from the whole MSI, while the denoising methods in~\cite{peng2014decomposable,xie2017kronecker} use different dictionaries for distinct groups.
\item Instead of enforcing the denoised groups to be exactly low-rank, we consider a more flexible model that each group is decomposed into a low-rank component and a non-low-rank component, as clean MSI groups are nearly low-rank in practice.
\end{itemize}

\par
A flowchart of the proposed MSI denoising method is shown in Figure~\ref{fig:fig1}. Full bands blocks of an MSI are extracted by window and then similar blocks are clustered into groups. Then we obtain 3-order tensor groups by unfolding each blocks at spectral domain. The three modes of a group corresponding to the spatial domain, the spectral domain and blocks. Shared overcomplete dictionaries of all groups in both the spatial mode and the spectral mode are learned via the proposed LTDL method, where a nearly low-rank structure is enforced for the tensor approximation. An effective algorithm based on the alternating direction method of multipliers (ADMM)~\cite{boyd2011distributed} is used to solve the optimization problem of the proposed LTDL method for MSI denoising. Experimental results demonstrate that the new method outperforms state-of-the-art methods in MSI denoising.

\par
The rest of the paper is organized as follows: Section~\ref{sec:work} reviews existing methods for image denoising based on the filter model, the sparsity model and the low-rank model. In section~\ref{sec:LTDL}, we present the proposed LTDL method for MSI denoising. Experimental results on synthetic data and MSIs are provided in section~\ref{sec:Ex}. At last, conclusions are given in section~\ref{sec:Conclusion}.
\section{Related Work}
\label{sec:work}

\begin{table}
\begin{center}
\begin{tabular}{|l|p{3.8cm}|l|}
\hline
Method & Model & Data \\
\hline\hline
 NL-means~\cite{buades2005non} & Filter & Matrix \\
 K-SVD~\cite{aharon2006k} & Sparsity & Matrix \\
 BM3D~\cite{dabov2007image} & Filter & Matrix \\
 CSR~\cite{dong2011sparsity} & Sparsity & Matrix \\
 WNNM~\cite{gu2014weighted} & Low Rank & Matrix \\
 LRTA ~\cite{renard2008denoising}& Low Rank & Tensor \\
 NLM3D~\cite{coupe2008optimized} & Filter & Tensor \\
 PARAFAC ~\cite{liu2012denoising}& Low Rank & Tensor \\
 BM4D~\cite{maggioni2013nonlocal} & Filter & Tensor \\
 Tdl~\cite{peng2014decomposable} & Sparsity and Low Rank & Tensor \\
 TenSR~\cite{qi2016tensr} & Sparsity & Tensor\\
 KBReg~\cite{xie2017kronecker} & Sparsity and Low Rank & Tensor \\
 the proposed LTDL & Sparsity and Low Rank & Tensor \\
\hline
\end{tabular}
\end{center}
\caption{A list of some recent denoising methods.}
\label{tab:t1}
\end{table}

To decompose images with noises, one require the priors of the different signals. The ``No Free Lunch'' theory in machine learning suggests that all algorithms perform the same for the randomised data and we can achieve good performance only when the data has some structure and some appropriate model is used. In the past decade, various model driven methods have been proposed for image denoising, and some of them are summarised in Table~\ref{tab:t1}.

\par
The first category of image denoising methods~\cite{buades2005non,dabov2007image,coupe2008optimized,maggioni2013nonlocal} uses different filters, e.g., the mean-value filter and the Wiener filter, to exploit the local correlation between adjacent pixels and/or the non-local correlation between similar small patches/blocks of an image. Another category of widely used methods considers the property that natural images or clustered groups of their patches usually exhibit low-rank structures. Different low-rank approximation methods have been proposed for image denoising such as nuclear norm regularization~\cite{dong2011sparsity}, Tucker low-rank decomposition~\cite{renard2008denoising} and CP low-rank decomposition~\cite{liu2012denoising}. Based on the sparsity model, dictionary learning methods assume that image patches/blocks are linear compositions of very few atoms selected from a dictionary. Dictionary learning and sparse representation in~\cite{aharon2006k,dong2011sparsity}, is firstly applied for 2D image denoising, and then extended to higher-order image denoising such as TenSR~\cite{qi2016tensr}. In recent years, researchers consider to exploit both the sparsity model and low rank model to better utilize the prior of the groups extracted from an MSI~\cite{peng2014decomposable,xie2017kronecker}. Unfortunately, in these recent methods, dictionaries (or called factors) are learned separately for each tensor group, which deviates from the principle of dictionary learning and significantly increases the total number of dictionary atoms. Our work differs with the existing methods for MSI denoising. Instead of using the exact low-rank model, we consider a nearly low-rank structure for each tensor group of an MSI, and learning dictionaries that are shared among all groups.

\section{Low-rank tensor dictionary learning for MSI denoising}
\label{sec:LTDL}
\subsection{Notation}

The following notations are used throughout this paper. The order of a tensor is the number of modes. Elements of an $N$-order tensor $\mathcal{X}\in\mathbb{R}^{I_1\times \ldots \times I_n\times \ldots\times I_N}$ are denoted by $x_{i_1 \ldots i_n \ldots i_N}$, where $i_n$ $(1\leq i_n\leq I_n)$ refers to the $n$th mode index. A mode-$n$ vector of an $N$-order tensor $\mathcal{X}$ is obtained from $\mathcal{X}$ by varying index in the $n$th mode while keeping the indices of other modes fixed.
The unfolding matrix of a tensor $\mathcal{X}$ at the $n$th mode is denoted as $X_{(n)}\in\mathbb{R}^{I_n\times (I_1 \ldots I_{n-1}I_{n+1}\ldots I_N)}$, where the columns are all mode-$n$ vectors of $\mathcal{X}$. The tensor $\mathcal{X}$ can be obtained by folding the matrix $X_{(n)}$ at the $n$th mode. The $n$-mode product of the tensor $\mathcal{X}$ and a matrix $U\in\mathbb{R}^{J\times I_n}$ is a tensor $\mathcal{X}\times _n U\in\mathbb{R}^{I_1\times \ldots \times I_{n-1}\times J\times I_{n+1}\times\ldots \times I_N}$, whose elements are computed by $(\mathcal{X}\times _n U)_{i_1 \ldots i_{n-1}ji_{n+1} \ldots i_N}=\sum_{i_n = 1}^{I_n}x_{i_1 \ldots i_N}u_{ji_n}$. The inner product of two tensors $\mathcal{X},\mathcal{Y}\in\mathbb{R}^{I_1\times I_2\times \ldots \times I_N}$ is the sum of the products of all entries, i.e., $\langle\mathcal{X},\mathcal{Y}\rangle = \sum_{i_1 =1}^{I_1} \ldots \sum_{i_N=1}^{I_N}x_{i_1 \ldots i_N}y_{i_1 \ldots i_N}.$
The $\ell_1$ norm and the Frobenius norm of a tensor are defined as
$\left \| \mathcal{X}\right \|  _1=\sum_{i_1 =1}^{I_1} \ldots \sum_{i_N=1}^{I_N}|x_{i_1 \ldots i_N}|$ and $\left \| \mathcal{X}\right \|  _F=(\sum_{i_1 =1}^{I_1}\ldots\sum_{i_N=1}^{I_N}x_{i_1 \ldots i_N}^2)^{1/2}$, respectively. The nuclear norm of a matrix is denoted by $\left \| X\right \|_*$, which is the sum of all singular values of $X$. The symbol of $\otimes$ denotes the Kronecker product of matrices.

\subsection{The observation of the nearly low-rank tensor structure in real MSI groups}
For an MSI $\mathcal{H}\in\mathbb{R}^{L\times W\times H}$ with $L\times W$ spatial size and $H$ spectral bands, we extract $S$ overlapping full-band blocks $\mathcal{H}^i\in\mathbb{R}^{d_L\times d_W\times H}$ by using a sampling window that traverses the whole MSI with step lengthes $p_L$ and $p_W$ in the two spacial coordinates. Each block $\mathcal{H}^i$ is unfolded in the spectral mode to be a matrix $H^{i}_{(3)}\in\mathbb{R}^{H \times d_Ld_W}$, which has a spatial mode and a spectral mode. To exploit the non-local self-similarity of images, blocks of the MSI can be clustered into $K$ groups, where each group forms a tensor (called a tensor group). The $k$th tensor group $\mathcal{X}^{(k)}\in\mathbb{R}^{d_L d_W\times H\times s^{(k)}}$ consists of $s^{(k)}$ similar blocks of the MSI and $\sum_{k=1}^{K}s^{(k)} = S$. Tensor groups are assumed with a low-rank structure in a variety of works ~\cite{peng2014decomposable,xie2016multispectral,xie2017kronecker}, owning to the correlations across the spatial domain, the spectral domain and similar blocks.

\par
To enforce the low-rank structure of the recovered signal from a noisy tensor group $\mathcal{X}^{(k)}$ of an MSI, there are at least two approaches. One approach is using the low-rank regularization. For example, by employing the convex tensor nuclear norm~\cite{liu2013tensor}, the tensor group denoising problem is casted into a convex optimization problem, which is given as
\begin{equation}
\min_{\hat{\mathcal{X}}^{(k)}} \frac{1}{2}\left \| \mathcal{X}^{(k)}- \hat{\mathcal{X}}^{(k)}\right \| ^2 _F + \sum_{i=1}^3 \beta_i \left \| \hat{X}_{(i)}^{(k)}\right \| _{*},
\label{3.2.1}
\end{equation}
where $\beta_i$ ($i=1,2,3$) are positive scalars that controls the low-rank penalties for different modes. The optimization problem in (\ref{3.2.1}) can be efficiently solved by the singular value thresholding~\cite{cai2010singular}. Another approach aims to find the rank-$(R_1^{(k)},R_2^{(k)},R_3^{(k)})$ approximation of the noisy tensor group $\mathcal{X}^{(k)}$ in an optimal least squares sense with $R_1^{(k)}<d_Ld_W$, $R_2^{(k)} <H$ and $R_3^{(k)} <s^{(k)}$ in ~\cite{peng2014decomposable}, which is given by
\begin{equation}
\min_{\mathcal{G}^{(k)},\{U_i^{(k)}\}}
\left \| \mathcal{X}^{(k)}\!-\!\mathcal{G}^{(k)}\!\times _1 \! U_1^{(k)}\!\times _2\! U_2^{(k)}\!\times _3\! U_3^{(k)}\right \| ^2 _F,
\label{3.2.2}
\end{equation}
where $\mathcal{G}^{(k)}\in\mathbb{R}^{R_1^{(k)}\times R_2^{(k)}\times R_3^{(k)}}$ is a smaller core tensor, and $U_i^{(k)}$ ($i=1,2,3$) are factors. The denoised tensor group is $\hat{\mathcal{X}} = \mathcal{G}^{(k)}\!\times _1 \! U_1^{(k)}\!\times _2\! U_2^{(k)}\!\times _3\! U_3^{(k)}$. The optimization problem in (\ref{3.2.2}) is nonconvex and different algorithms are proposed in literature such as higher-order singular value decomposition (HOSVD)~\cite{de2000multilinear} and higher-order orthogonal iteration (HOOI)~\cite{de2000best}.

\begin{figure}[t!]
\setlength{\abovecaptionskip}{0.5cm}
\setlength{\belowcaptionskip}{-0.5cm}
\centering
\subfigure[]{
\label{fig:lr(a)}
\includegraphics[width=0.37\columnwidth]{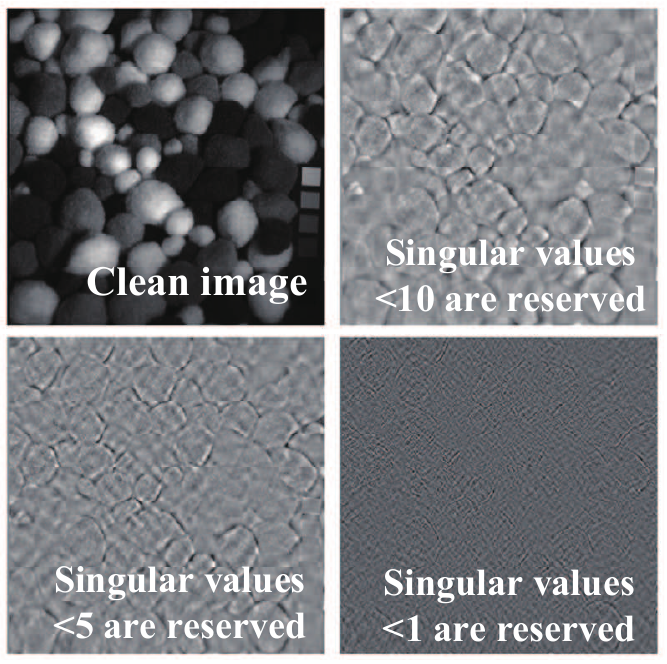}}
\subfigure[]{
\label{fig:lr(b)}
\includegraphics[width=0.57\columnwidth]{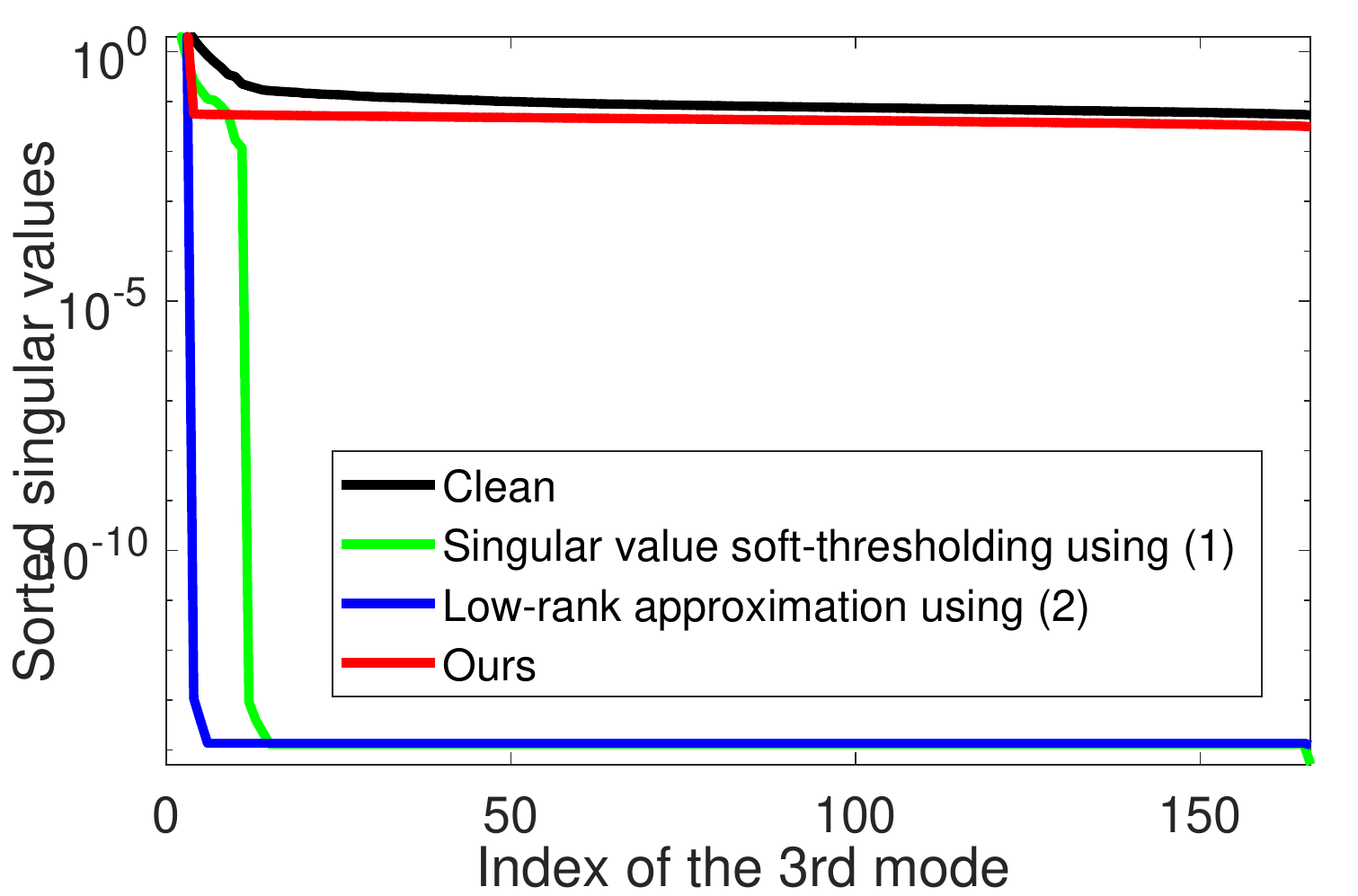}}
\label{fig:lr}
\caption{Investigation on the distribution of singular values of real MSIs and the effectiveness of different methods. (a) The clean ``Pompoms'' MSI and different ``Pompoms'' MSIs with large singular values removed. (b) The curves for the sorted singular values of a tensor group in the 3rd mode of different low-rank based methods.}
\end{figure}
\par
However, we observe that usually tensor groups of clean real MSIs are not exactly low-rank. As shown in Fig.~\ref{fig:lr(b)}, the curve (in black) for the sorted singular values of a tensor group of the clean ``pompoms'' MSI in the 3rd mode has a very long tail, i.e., many singular values are close to $0$ rather than exactly $0$. As shown in ~\ref{fig:lr(a)}, these small singular values contain the texture information of the MSI. Therefore, enforcing an exactly low-rank structure of a tensor group would lead to the lost of important information of the MSI. To avoid this drawback, we consider a nearly low-rank structure for each tensor group and pose the following optimization problem for denoising:
\begin{equation}
\begin{split}
&\min_{\mbox{\tiny$\hat{\mathcal{X}}^{(k)},\mathcal{G}^{(k)},\{U_i^{(k)}\}$}}\left \| \mathcal{X}^{(k)}-\hat{\mathcal{X}}^{(k)}\right \| ^2 _F\\
&\qquad\  \!+\!\lambda_r \left \| \hat{\mathcal{X}}^{(k)}-\!\mathcal{G}^{(k)}\!\times _1 \! U_1^{(k)}\!\times _2\! U_2^{(k)}\!\times _3\! U_3^{(k)}\right \| ^2 _F,
\end{split}
\label{3.2.3}
\end{equation}
where $\hat{\mathcal{X}}^{(k)}$ is the $k$th recovered tensor group, and $\lambda_r$ is the weight to balance fidelity and the low-rank structure. Note that the small singular values of $\hat{\mathcal{X}}^{(k)}$ results from $\hat{\mathcal{X}}^{(k)}-\!\mathcal{G}^{(k)}\!\times _1 \! U_1^{(k)}\!\times _2\! U_2^{(k)}\!\times _3\! U_3^{(k)}$, and are penalized via the Frobenius norm. The optimization problem in (\ref{3.2.3}) can be solved by updating the variables in $\hat{\mathcal{X}}^{(k)}$ and $\mathcal{T}^{(k)} = \mathcal{G}^{(k)}\!\times _1 \! U_1^{(k)}\!\times _2\! U_2^{(k)}\!\times _3\! U_3^{(k)}$ alternately. As shown in Fig.~\ref{fig:lr(b)}, with appropriate choose of the weight $\lambda_r$, the curve for the sorted singular values of the proposed method is close to the curve of the clean MSI tensor group, while the methods corresponding to (\ref{3.2.1}) and (\ref{3.2.2}) lose information embedded in the small singular values.


\subsection{The proposed LTDL method for MSI denoising}
\label{sec:LTDL model}
In this subsection, we introduce the proposed MSI denoising method that learns shared tensor overcomplete dictionaries and considers the nearly low-rank structure. Two dictionaries, i.e., $D^{a}$ and $D^{e}$, are learned from all tensor groups of an MSI, where $D^{a}$ corresponds to the spatial domain and $D^{e}$ corresponds to the spectral domain. We define $\tau _{a}$ and $\tau _{e}$ as the redundancy ratios (i.e., the ratio of the number of columns to the number of rows, $\tau _{a}\geq 1$ and $\tau _{e}\geq 1$) corresponding to the spatial dictionary and the spectral dictionary, respectively. The proposed LTDL method for MSI denoising is formulated as follows:
\begin{equation}
\begin{split}
&\min_{\mbox{\tiny$\begin{array}{c}
D^{a},D^{e}\\
\{\mathcal{Z}^{(k)},\mathcal{G}^{(k)}\\
U_i^{(k)}\}\\
\end{array}$}}
\!\mbox{\small$\displaystyle{\sum_{k=1}^K}\left(\left \| \mathcal{X}^{(k)}\!-\!\mathcal{Z}^{(k)}\!\times _1\! D^{a}\!\times _2\! D^{e}\right \| ^2 _F+ \lambda _s \left \|\mathcal{Z}^{(k)}\right \| _1\right.$}\qquad\qquad
\\
&\quad\mbox{\small$\left.+ \lambda _r \left \| \mathcal{Z}^{(k)}\!\times _1 \! D^{a}\!\times _2\! D^{e}\! - \!\mathcal{G}^{(k)}\!\times _1 \! U_1^{(k)}\!\times _2\! U_2^{(k)}\!\times _3\! U_3^{(k)}\right \| ^2 _F \right)$}
\\
&\quad\quad \text{s.t.} \quad \left \| D^{a}(:,r)\right \| ^2 _2 =1\ \text{for}\  r=1,\ldots, \tau _{a}d_L d_W
\\
&\quad\qquad\quad\,\left \| D^{e}(:,r)\right \| ^2 _2 =1\ \text{for}\  r=1,\ldots, \tau _{e}H,
\end{split}
\label{LTDL}
\end{equation}
where $\mathcal{Z}^{(k)}\in\mathbb{R}^{\tau_{a}d_Ld_W\times \tau_{e}H\times s^{(k)}}$ is the representation of $k$th tensor group, $\lambda_s$ and $\lambda_r$ are the weights corresponding to regularization for sparsity and low-rank, respectively. For the last term of the object in (\ref{LTDL}), $\mathcal{T}^{(k)}=\mathcal{G}^{(k)}\times _1U_1^{(k)}\times _2 U_2^{(k)}\times _3 U_3^{(k)}$ denotes the rank-$(R_1^{(k)},R_2^{(k)},R_3^{(k)})$ approximation of the reconstructed tensor group $\mathcal{Z}^{(k)}\times _1 D^{a}\times _2 D^{e}$.

\subsection{Algorithm development}
The proposed optimization problem in (\ref{LTDL}) is nonconvex. Here, we employ the alternating direction method of multipilers(ADMM)~\cite{boyd2011distributed}, which leads to solve several subproblems.

\par
We first introduce $K$ auxiliary tensors $\mathcal{C}^{(k)}$($k=1,\ldots,K$) and the optimization problem in (\ref{LTDL}) is equivalent to
\begin{equation}
\begin{split}
\min_{\mbox{\tiny$\begin{array}{c}
D^{a},D^{e}\\
\{\mathcal{Z}^{(k)},\mathcal{C}^{(k)}\\
\mathcal{T}^{(k)}\}\\
\end{array}$}}
\!&\!\mbox{\small$\displaystyle{\sum_{k=1}^K}\left(\left \| \mathcal{X}^{(k)}-\mathcal{Z}^{(k)}\times _1 D^{a}\times _2 D^{e}\right \| ^2 _F+ \lambda _s \left \|\mathcal{C}^{(k)}\right \| _1\right.$}\qquad\qquad
\\
&\qquad\mbox{\small$\left.+ \lambda _r \left \| \mathcal{Z}^{(k)}\times _1 D^{a}\times _2 D^{e} - \mathcal{T}^{(k)}\right \| ^2 _F \right)$}
\\
\text{s.t.}\qquad &\left \| D^{a}(:,r)\right \| ^2 _2 =1\ \text{for}\  r=1,\ldots, \tau _{a}d_L d_W
\\
&\left \| D^{e}(:,r)\right \| ^2 _2 =1\ \text{for}\  r=1,\ldots, \tau _{e}H
\\
&\ \mathcal{C}^{(k)} = \mathcal{Z}^{(k)}\ \text{for}\  k=1,\ldots,K,
\end{split}
\label{LTDL-ADMM}
\end{equation}
The augmented Lagrangian function for the above problem can be given as:
\begin{equation}
\begin{split}
\mathcal{L}_{\rho}&\mbox{\small$\left( D^{a},D^{e},\{\mathcal{Z}^{(k)},
\mathcal{C}^{(k)},\mathcal{T}^{(k)},
\mathcal{Y}^{(k)}\}\right)$}
\\
&\mbox{\small$=\displaystyle{\sum_{k=1}^K} \left(\left \| \mathcal{X}^{(k)}-\mathcal{Z}^{(k)}\times _1 D^{a}\times _2 D^{e}\right \| ^2 _F\right.$}
\\
&\mbox{\small$+\lambda _s \left \|\mathcal{C}^{(k)}\right \| _1 +\langle \mathcal{C}^{(k)}-\mathcal{Z}^{(k)}, \mathcal{Y}^{(k)}\rangle +\dfrac{\rho}{2}\left \| \mathcal{C}^{(k)}-\mathcal{Z}^{(k)}\right \|^2_F$}
\\
&\mbox{\small$\left.+\lambda_r \left \| \mathcal{Z}^{(k)}\times _1 D^{a}\times _2 D^{e}-\mathcal{T}^{(k)}\right\|^2_F\right)$},
\end{split}
\label{LTDL-L}
\end{equation}
where $\mathcal{Y}^{(k)} (k=1,\ldots,K)$ are the Lagrange multipiers, $\rho$ is a positive scalar, and columns of the dictionaries $D^{a}$ and $D^{e}$ are constrained by a unit power.

\par
Now we introduce the strategy for solving (\ref{LTDL-ADMM}) based on the ADMM. To minimize the augmented Lagrange function (\ref{LTDL-L}), we fix the dictionaries $D^{a}$ and $D^{e}$, and all multipliers $\mathcal{Y}^{(k)}$. Then (\ref{LTDL-L}) can be split into $K$ separate optimization problems with objects as:
\begin{equation}
\begin{split}
&\mathcal{L}^{(k)}_{\rho}\mbox{\small$\left( \mathcal{Z}^{(k)},
\mathcal{C}^{(k)},\mathcal{T}^{(k)}\right)\!=\!\left \| \mathcal{X}^{(k)}-\mathcal{Z}^{(k)}\times _1 D^{a}\times _2 D^{e}\right \| ^2 _F$}
\\
&\quad\mbox{\small$+\lambda_s\left\|\mathcal{C}^{(k)}\right \| _1 +\langle \mathcal{C}^{(k)}-\mathcal{Z}^{(k)}, \mathcal{Y}^{(k)}\rangle +\dfrac{\rho}{2}\left \| \mathcal{C}^{(k)}-\mathcal{Z}^{(k)}\right\|^2_F$}
\\
&\quad\mbox{\small$+\lambda_r\left\|\mathcal{Z}^{(k)}\times _1 D^{a}\times _2 D^{e} - \mathcal{T}^{(k)}\right \|^2 _F$}.
\end{split}
\label{LTDL-Lk}
\end{equation}
$\mathcal{T}^{(k)}$, $\mathcal{Z}^{(k)}$ and $\mathcal{C}^{(k)}$ can be updated alternatively by solving the optimization problem with all the other variables fixed. In specific, to update $\mathcal{T}^{(k)}$, one needs to solve the tensor low multilinear rank approximation problem given by
\begin{equation}
\min_{\mathcal{T}^{(k)}}
\left \| \mathcal{Z}^{(k)}\times _1 D^{a}\times _2 D^{e} - \mathcal{T}^{(k)}\right \| ^2 _F,
\label{CosF-T}
\end{equation}
which can be handled by HOOI~\cite{de2000best} and the solution is denoted as
\begin{equation}
\mathcal{T}^{(k)} =\text{HOOI}\left(\mathcal{Z}^{(k)}\times _1 D^{a}\times _2 D^{e},R_1^{(k)},R_2^{(k)},R_3^{(k)}\right).
\label{Sol-T}
\end{equation}
To update $\mathcal{Z}^{(k)}$, with irrelevant terms removed, the problem of $\min_{\mathcal{Z}^{(k)}}\mathcal{L}^{(k)}_{\rho}$ becomes:
\begin{equation}
\begin{split}
&\min_{\mathcal{Z}^{(k)}}\mbox{\small$\left \| \mathcal{X}^{(k)}-\mathcal{Z}^{(k)}\times _1 D^{a}\times _2 D^{e}\right \| ^2 _F+\!\langle \mathcal{C}^{(k)}-\mathcal{Z}^{(k)}, \mathcal{Y}^{(k)}\rangle $}
\\
&\ \mbox{\small$+\dfrac{\rho}{2}\left \| \mathcal{C}^{(k)}-\mathcal{Z}^{(k)}\right\|^2_F+\!\lambda_r \left \| \mathcal{Z}^{(k)}\times_1D^{a}\times_2 D^{e}- \mathcal{T}^{(k)}\right\|^2 _F$},
\label{CosF-Z}
\end{split}
\end{equation}
which has a closed-form solution
\begin{equation}
\begin{split}
Z_{(3)}^{(k)}= &\left(\left(2X_{(3)}^{(k)}+2\lambda_r T_{(3)}^{(k)}\right)D+\rho C_{(3)}^{(k)}+Y_{(3)}^{(k)}\right)\\
&\qquad \left(\left(2+2\lambda_r\right)D^T D+\rho I\right)^{-1},
\end{split}
\label{Sol-Z}
\end{equation}
where $D=D^{e}\otimes D^{a}$ and $I$ is an identity matrix. The tensor $\mathcal{Z}^{(k)}$ can be
obtained by folding $Z_{(3)}^{(k)}$ at the $3$rd mode. Updating $\mathcal{C}^{(k)}$ requires to solve the following optimization problem:
\begin{equation}
\begin{split}
\min_{\mathcal{C}^{(k)}}\ \lambda _s \left \|\mathcal{C}^{(k)}\right \| _1 +\dfrac{\rho}{2}\left \| \mathcal{C}^{(k)}-\left(\mathcal{Z}^{(k)}-\dfrac{1}{\rho}\mathcal{Y}^{(k)}
\right)\right \| ^2 _F,
\end{split}
\label{CosF-C}
\end{equation}
which leads to the solution
\begin{equation}
\begin{split}
\mathcal{C}^{(k)}\!=\!\text{soft}_{\frac{\lambda_s}{\rho}}(\mathcal{M}^{(k)}),
\end{split}
\label{Sol-C}
\end{equation}
where $\text{soft}(\cdot)$ denotes the soft-thresholding operator and $\mathcal{M}^{(k)}\!=\!\mathcal{Z}^{(k)}\!-\!\mathcal{Y}^{(k)}/\rho$ . The elements of $\mathcal{C}^{(k)}$ is $c^{(k)}_{ijk} \!=\! \text{sign}(m^{(k)}_{ijk})\text{max}\{0,\left| m^{(k)}_{ijk}\right|-\dfrac{\lambda_s}{\rho}\}$, where $\text{sign}(\cdot)$ denotes the sign function. Note that variables for different tensor groups ($k=1,\ldots,K$) can be updated in parallel.

\par
Next we consider to update the dictionaries $D^{a}$ and $D^{e}$, which are shared by all groups. By letting $\mathcal{O}^{(k)}=\frac{\mathcal{X}^{(k)}+\lambda_r\mathcal{T}^{(k)}}{1+\lambda_r}$, the optimization problem turns to be
\begin{equation}
\begin{split}
\min_{D^{a},D^{e}}&\left \| \mathcal{O}-\mathcal{Z}\times _1 D^{a}\times _2 D^{e}\right \| ^2 _F ,
\\
\text{s.t.}\quad&\left \| D^{a}(:,r)\right \| ^2 _2 =1\ \text{for}\  r=1,\ldots,\tau _{a}d_L d_W
\\
&\left \| D^{e}(:,r)\right \| ^2 _2 =1\ \text{for}\  r=1,\ldots, \tau _{e}H
\end{split}
\label{CosF-D}
\end{equation}
where $\mathcal{O}=\left(\mathcal{O}^{(1)},\ldots,\mathcal{O}^{(K)}\right)$ and $\mathcal{Z}=\left(\mathcal{Z}^{(1)},\ldots,\mathcal{Z}^{(K)}\right)$ are obtained by stacking all groups of $\mathcal{O}^{(k)}$ and $\mathcal{Z}^{(k)}$ at the $3$rd mode, respectively. Define $\mathcal{A}=\mathcal{Z}\times _2 D^{e}$. Then $D^{a}$ can be updated by
\begin{equation}
\begin{split}
\min_{D^{a}}\quad&\left \| O_{(1)}-D^{a}A_{(1)}\right \| ^2 _F
\\
\text{s.t.}\quad&\left \| D^{a}(:,r)\right \| ^2 _2 =1\ \text{for}\  r=1,\ldots,\tau _{a}d_L d_W,
\end{split}
\label{CosF-Dspa}
\end{equation}
where $O_{(1)}\in\mathbb{R}^{d_Ld_W\times HS}$ and $A_{(1)}\in\mathbb{R}^{\tau_{a}d_Ld_W\times HS}$. The optimization problem in (\ref{CosF-Dspa}) is a quadratically constrained quadratic programming problem and can be solved using a Lagrange dual~\cite{lee2007efficient}. Thus, the spatial dictionary $D^{a}$ can be updated by
\begin{equation}
{D^{a}}=\left(O_{(1)}A_{(1)}^T\right)\left(A_{(1)}A_{(1)}^T+\Gamma\right)^{-1},
\label{Sol-Da}
\end{equation}
where $\Gamma=\text{diag}([\gamma_1,\ldots,\gamma_{\tau_{a}d_Ld_W}])$ and $\gamma_r$ ($r=1,\ldots,\tau _{a}d_L d_W$) are dual variables whose values are obtained by solving the dual problem. Similarly, letting $\mathcal{E}=\mathcal{Z}\times _1 D^{a}$, $D^{e}$ can be obtained by solving the following problem:
\begin{equation}
\begin{split}
\min_{D^{e}}\quad& \left \| O_{(2)}-D^{e}E_{(2)}\right \| ^2 _F
\\
\text{s.t.}\quad&\left \| D^{e}(:,r)\right \| ^2 _2 =1\ \text{for}\  r=1,\ldots, \tau _{e}H.
\end{split}
\label{CosF-De}
\end{equation}
Then the spectral dictionary $D^{e}$ is updated by
\begin{equation}
{D^{e}}=\left(O_{(2)}E_{(2)}^T\right)\left(E_{(2)}E_{(2)}^T+\Delta\right)^{-1},
\label{Sol-De}
\end{equation}
where $\Delta=\text{diag}([\delta_1,\ldots,\delta_{\tau _{e}H}])$ and $\delta_r$ ($r=1,\ldots, \tau _{e}H$) are optimized dual variables of (\ref{CosF-De}).

\par
For the last step, we update the $K$ Lagrange multipliers in turn by
\begin{equation}
\mathcal{Y}^{(k)}:=\mathcal{Y}^{(k)}+\rho \left(\mathcal{C}^{(k)}-\mathcal{Z}^{(k)}\right).
\label{Up-Y}
\end{equation}
The proposed LTDL method for MSI denoising is summarized in Algorithm~\ref{alg:1}.

\begin{algorithm}[htbp]
\renewcommand{\algorithmicrequire}{\textbf{Input:}}
\renewcommand\algorithmicensure {\textbf{Output:} }
\caption{Algorithm for MSI denoising}
\label{alg:1}
\begin{algorithmic}[1]
\REQUIRE the MSI of $\mathcal{H}\in\mathbb{R}^{L\times W\times H}$
\ENSURE  denoised MSI $\mathcal{H}_{de}$, spatial dictionary $D^{a}$ and spectral dictionary $D^{e}$
\label{alg:1}
\STATE Construct groups $\mathcal{X}^{(k)}(k=1,\ldots,K)$ by the extracting, clustering and unfolding process of the $\mathcal{H}$
\STATE  Initialize $D^{a},D^{e},\left\{\mathcal{C}^{(k)},\mathcal{Z}^{(k)},\mathcal{Y}^{(k)}\right\}_{k=1}^K$
\WHILE{not converged}
\FOR {$k=1:K$}
\STATE{update $\mathcal{T}^{(k)}$ by (\ref{Sol-T})}
\STATE{update $\mathcal{Z}^{(k)}$ via folding $Z_{(3)}^{(k)}$ by (\ref{Sol-Z})}
\STATE{update $\mathcal{C}^{(k)}$ by (\ref{Sol-C})}
\ENDFOR
\STATE{update $D^{a}$  by (\ref{Sol-Da})}
\STATE{update $D^{e}$  by (\ref{Sol-De})}
\FOR {$k=1:K$}
\STATE{update $\mathcal{Y}^{(k)}$ by (\ref{Up-Y})}
\ENDFOR
\STATE{$\rho:=\mu\rho$}
\ENDWHILE
\STATE{Reconstruct groups $\hat{\mathcal{X}}^{(k)}\!=\!\mathcal{Z}^{(k)}\!\times_1\! D^{a}\!\times_2\!D^{e} (k=1,\ldots,K)$}
\STATE{Aggregate $\hat{\mathcal{X}}^{(k)}$ to form the denoised MSI $\mathcal{H}_{de}$}
\end{algorithmic}
\end{algorithm}

In the following theorem, we provide a weak convergence condition for the proposed algorithm, where we assume the dictionaries are fixed. By alternatively updating the dictionaries and the other variables, it can be guaranteed that the value of the cost function in (\ref{LTDL}) would not increase. In practice, to speed up the algorithm, one can update the dictionaries in each iteration without waiting for the other variables achieving convergence. Although we cannot provide formal convergence guarantees in this case, we do not have convergence problems in our experiments.

$\mathbf{Theorem\ 1.}$ When $D^a$ and $D^e$ are fixed, for $l$th iteration, the sequences $\{\mathcal{Z}_l^{(k)}\}$,  $\{\mathcal{C}_l^{(k)}\}$ and $\{\hat{\mathcal{X}}_l^{(k)}\}$ ($k = 1,\ldots ,K$) satisfy:
\begin{equation}
\begin{split}
&\left\|\mathcal{Z}_l^{(k)}-\mathcal{C}_l^{(k)}\right\|_F\leq O\left(\frac{1}{\mu^{l-1}\rho_0}\right)
\\
&\left\|\mathcal{Z}_{l+1}^{(k)}-\mathcal{Z}_l^{(k)}\right\|_F\leq O\left(\frac{1}{\mu^{l-1}\rho_0}\right)\\
&\left\|\hat{\mathcal{X}}_{l+1}^{(k)}-\hat{\mathcal{X}}_l^{(k)}\right\|_F\leq O\left(\frac{1}{\mu^{l-1}\rho_0}\right).
\end{split}
\label{Conv-ZC}
\end{equation}
where $\rho_0$ is initial value. Proofs are provided in the supplementary document.
\section{Experimental results}
\label{sec:Ex}

In this section, we evaluate the effectiveness of the proposed LTDL using both synthetic data and real MSIs.

\begin{figure}[!tb]
\begin{center}
\includegraphics[width=0.9\linewidth]{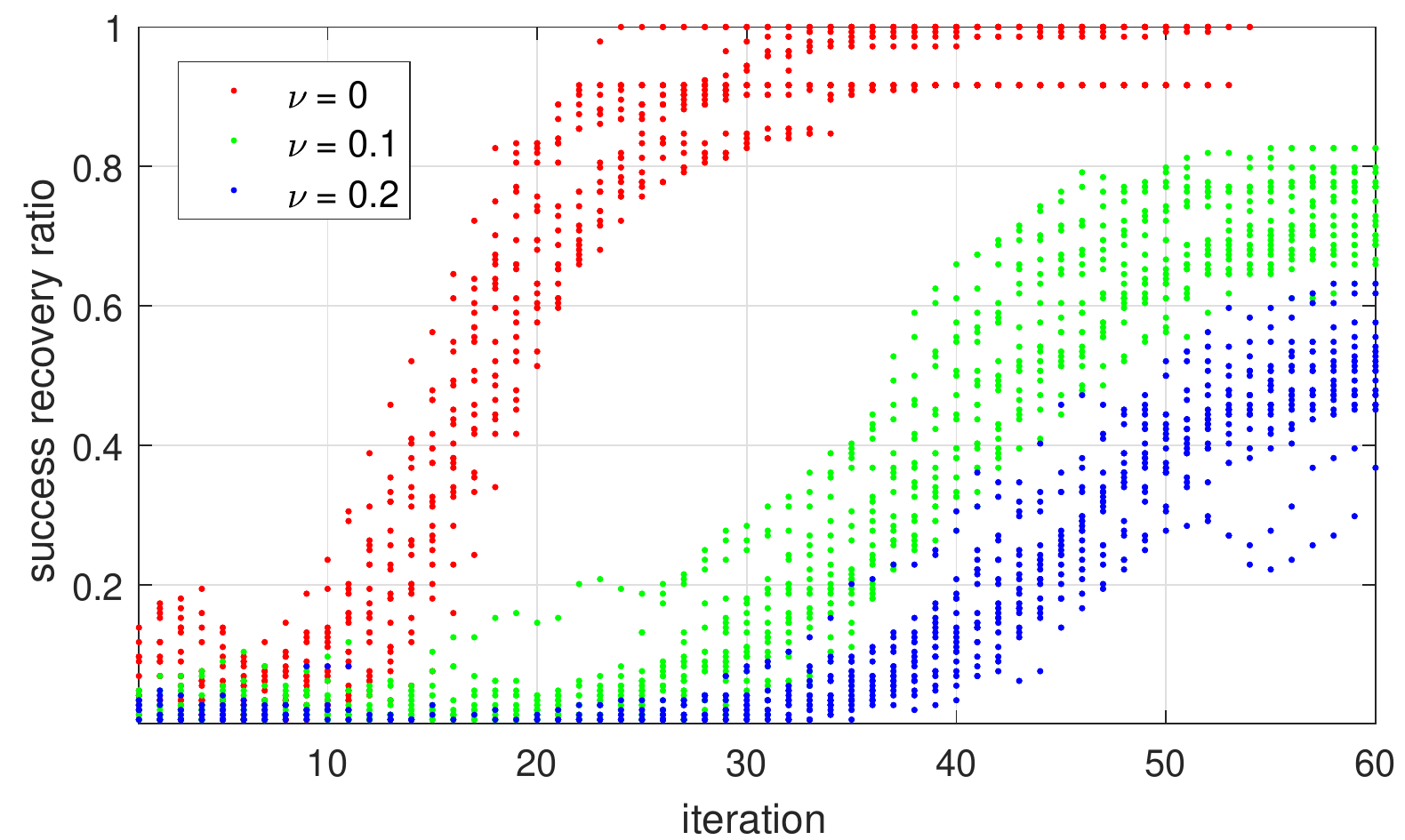}
\end{center}
   \caption{Success recovery ratio of the ground truth dictionary.}
\label{fig:f3}
\end{figure}
\subsection{Dictionary Learning Performance with Synthetic Data}
We first evaluate the dictionary learning performance of the proposed method on synthetic data. Both the spatial dictionary $D^a\in\mathbb{R}^{10\times 12}$ and the spectral dictionary $D^e\in\mathbb{R}^{10\times 12}$ are generated randomly with normalized columns. For each tensor group, its tensor representation $\mathcal{Z}^{(k)}\in\mathbb{R}^{12\times 12\times 12}$ is generated randomly with the sparsity level 6, i.e., no more than 6 dictionary atoms are used to constitute the tensor group. The nonzero components in $\mathcal{Z}^{(k)}$ can be rewritten as a $12\times 6$ matrix, which is generated as the production of two random matrices $G_k\in\mathbb{R}^{12\times 2}$ and $F_k\in\mathbb{R}^{2\times 5}$. Therefore, all generated tensor groups have ranks no higher than $(6,6,2)$. We generate 200 different tensor groups in this way, and then add Gaussian noises with standard deviation $\nu$. The learned dictionaries are compared against all atoms of the generated dictionaries and we find the most close pair via $1-|d_i^T\hat{d}_i|$, where $d_i$ is a generated dictionary atom and $\hat{d}_i$ is a recovered dictionary atom\footnote{We consider the equivalent dictionary $D^e\otimes D^a$ in the experiments.}. The learning is seemed as success if the distance is less than $0.1$. The convergence behaviour of the proposed algorithm, i.e., success recovery ratio of the ground truth dictionary versus the number of iterations is shown in Figure \ref{fig:f3}, where we provide the results of 20 trials for each noise level. With the decrease of the noise variance, the performance of success recovery ratio is improved.


\subsection{Denoising Performance for Real MSIs}
\begin{figure*}[t!]
\begin{center}
\includegraphics[width=0.95\linewidth]{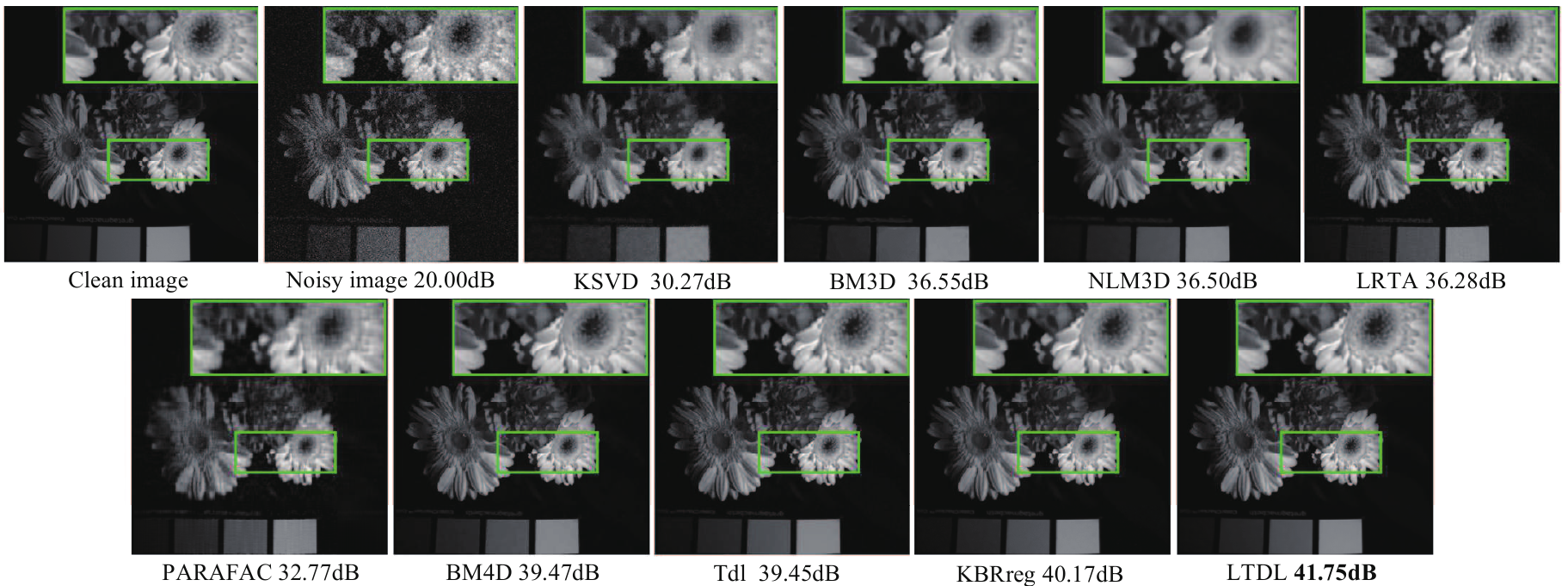}
\end{center}
   \caption{Denoising results of the 590nm band image of the ``flowers'' MSI ($\nu = 0.1$).}
\label{fig:MSI}
\end{figure*}

\renewcommand{\arraystretch}{1.5} %
\begin{table*}[ht]

  \centering
  \fontsize{6.5}{8}\selectfont
  \caption{MSI denoising performance on the Columbia dataset.}
  \label{tab:performance_comparison}
    \begin{tabular}{|c|c|c|c|c|c|c|c|c|}
    \hline
    \multirow{2}{*}{Method}&
    \multicolumn{4}{c|}{$\nu=0.1$}&\multicolumn{4}{c|}{ $\nu=0.2$}\cr\cline{2-9}
    &PSNR&SSIM&SAM&ERGAS&PSNR&SSIM&SAM&ERGAS\cr
    \hline
    Noisy image
    &{20.00$\pm$0.00}&{0.14$\pm$0.07}&0.94$\pm$0.24&552.89$\pm$159.26
    &13.98$\pm$0.00&0.05$\pm$0.03&1.13$\pm$0.20&1105.74$\pm$318.50
    \cr\hline
    K-SVD
    &{30.07$\pm$1.30}&{0.61$\pm$0.04}&{0.53$\pm$0.22} & 170.02$\pm$38.26
    &{27.30$\pm$1.44}&{0.46$\pm$0.05}&{0.61$\pm$0.23} & 233.87$\pm$51.99
    \cr\hline
    BM3D
    &36.92$\pm$2.88&0.92$\pm$0.03&0.21$\pm$0.09 & 78.43$\pm$23.40
    &33.39$\pm$2.93&0.86$\pm$0.06&0.28$\pm$0.11 & 117.97$\pm$36.64
    \cr\hline
    NLM3D
    &36.47$\pm$2.93&0.93$\pm$0.04&0.24$\pm$0.10 & 85.64$\pm$21.54
    &33.44$\pm$2.83&0.87$\pm$0.05&0.33$\pm$0.15 & 119.53$\pm$30.15
    \cr\hline
    LRTA
    &36.58$\pm$2.78&0.89$\pm$0.05&0.22$\pm$0.11 & 82.09$\pm$26.02
    &33.08$\pm$2.68&0.82$\pm$0.08&0.28$\pm$0.12 & 122.36$\pm$37.41
    \cr\hline
    PARAFAC
    &33.16$\pm$4.17&0.84$\pm$0.11&0.28$\pm$0.14 & 127.31$\pm$65.73
    &31.10$\pm$3.00&0.73$\pm$0.08&0.42$\pm$0.20 & 155.92$\pm$61.05
    \cr\hline
    BM4D
    &39.49$\pm$2.29&0.94$\pm$0.02&0.22$\pm$0.12 & 57.81$\pm$12.63
    &35.53$\pm$2.10&0.87$\pm$0.03&0.34$\pm$0.17 & 91.37$\pm$19.21
    \cr\hline
    Tdl
    &39.22$\pm$2.37&0.94$\pm$0.02&0.17$\pm$0.10 & 59.67$\pm$14.28
    &35.12$\pm$2.05&0.87$\pm$0.03&0.27$\pm$0.15 & 95.33$\pm$22.70
    \cr\hline
    KBRreg
    &40.63$\pm$2.15&0.95$\pm$0.03&0.24$\pm$0.22 & 51.80$\pm$13.47
    &37.57$\pm$2.51&0.92$\pm$0.05&0.25$\pm$0.24 & 72.98$\pm$20.00
    \cr\hline
    {\bf Ours}
    &{\bf 41.50}$\pm$2.82&{\bf 0.97}$\pm$0.01&{\bf 0.09}$\pm$0.03 & {\bf 46.18}$\pm$12.84
    &{\bf 38.13}$\pm$ 2.73&{\bf 0.95}$\pm$0.02&{\bf 0.13}$\pm$ 0.07 & {\bf 67.65}$\pm$ 18.40
    \cr\hline
    \end{tabular}
\end{table*}

\textbf{MSI datasets}: Two real MSI datasets, i.e., the Columbia dataset~\cite{yasuma2010generalized}\footnote {http://www.cs.columbia.edu/CAVE/databases/multispectral/} and the urban area HYDICE MSI\footnote {https://erdc-library.erdc.dren.mil/xmlui/handle/11681/2925}, are used to evaluate the denoising performance of the proposed algorithm. The Columbia dataset includes $32$ scenes that are separated into 5 sections. Each MSI has the size of $512\times512$ in space and includes full spectral resolution reflectance data from $400nm$ to $700nm$ at $10nm$ steps, which leads to 31 bands. These MSIs are adopted as clean data, and we generate additive noises to evaluate the denoising performance of the proposed algorithm. For the natural urban area HYDICE MSI, we select the bands from $178$ to $207$, which are severely damaged (i.e., the noise is not generated).

\textbf{Experimental settings}: In the proposed LTDL method, the spatial window size is set as $d_L=d_W=7$ and the step size is set as $p_L=p_W=3$. Groups of an MSI are clustered by using the k-means++~\cite{arthur2007k}, and each cluster forms a tensor group. We set $\rho=0.01$ and $\mu=1.3$ as the parameters in the ADMM of our algorithm. The redundancy ratio of dictionaries are set as $\tau_a = \tau_e  = 1.5$. We add Gaussian noise with the mean $0$ and the standard deviation $\nu$ on the whole MSIs in the Columbia dataset. We set the sparsity weight $\lambda_s = 0.1\nu$ and the low-rank weight $\lambda_r = 500\nu$. The rank parameters $\{R_1^{(k)},R_2^{(k)},R_3^{(k)}\}^K$ in (\ref{Sol-T}) are estimated by using the tensor rank estimation method proposed in~\cite{yokota2017robust}. For unknown noises, we suggest that the $(R_1^{(k)},R_2^{(k)},R_3^{(k)})$ is smaller than the size of $k$th tensor group.

\par
The proposed algorithm is compared with both denoising methods for 2D images including K-SVD~\cite{aharon2006k} and BM3D~\cite{dabov2007image} and denoising methods for 3D images including NLM3D~\cite{coupe2008optimized}, LRTA ~\cite{renard2008denoising}, PARAFAC ~\cite{liu2012denoising}, BM4D~\cite{maggioni2013nonlocal}, Tdl~\cite{peng2014decomposable}, and KBRreg~\cite{xie2017kronecker}. For denoising methods for 2D images, each MSI is processed band by band as multiple 2D images.
\begin{figure*}[ht]
\begin{center}
\includegraphics[width=0.95\linewidth]{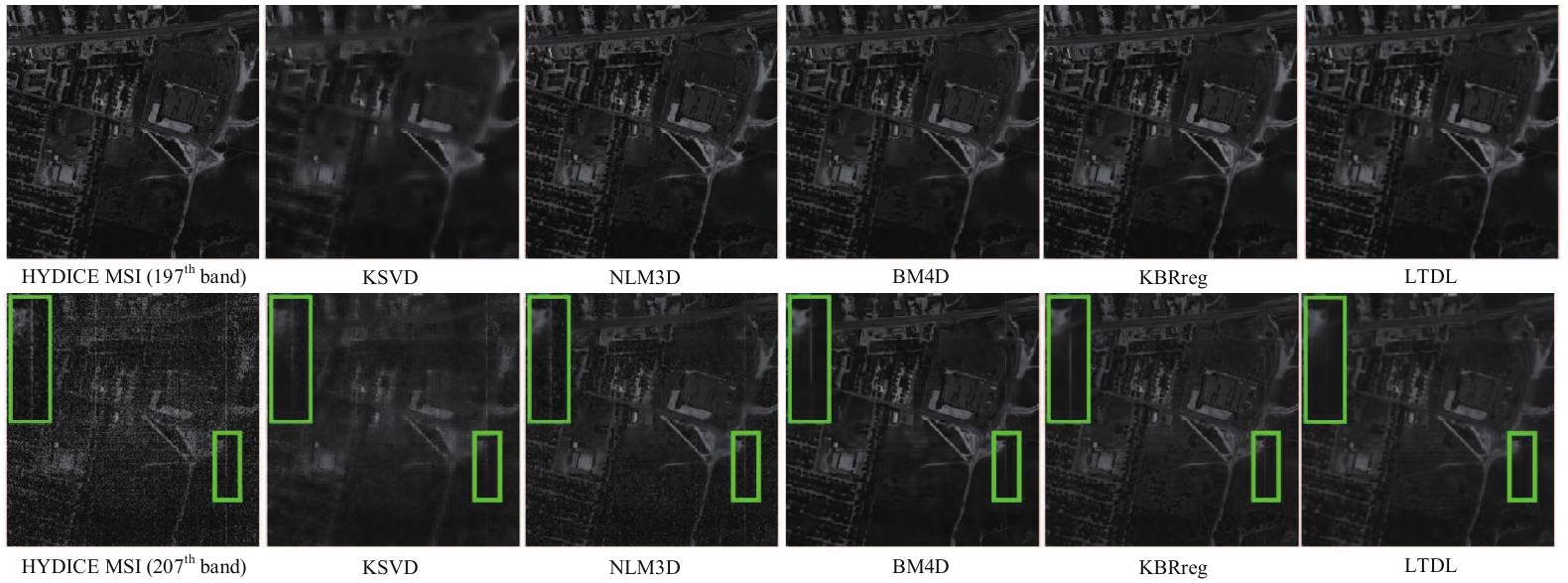}
\end{center}
   \caption{Denoising results for the HYDICE MSI (the 197th band on the top and the 207th band on the bottom).}
\label{fig:hsi}
\end{figure*}

\textbf{MSI denoising with generated noises}: We generate Gaussian noises for MSIs in the Columbia dataset, and employ four different performance indicators including peak signal-to-noise ratio (PSNR), structural similarity (SSIM), spectral angle mapper (SAM)~\cite{yuhas1993determination} and dimensionless global relative error of synthesis (ERGAS)~\cite{wald2002data}, which are widely used to evaluate recovery quality for MSIs. Recovered MSIs with higher PSNR and SSIM or lower SAM and ERGAS are considered with better quality. PSNR and SSIM are classical spatial-based quality indices, while ERGAS and SAM are spectral-based quality indices. We report the averaged denoising performance and the standard deviation of all the compared methods in Table \ref{tab:performance_comparison}. It can be observed that the proposed method outperforms all the competing methods under all of the four different quality indices.

\par
To visualize the denoising performance, we display the denoised 590nm band image of the ``flowers'' MSI in Figure~\ref{fig:MSI}. By zooming in the petal part of the image, it can be observed that the proposed LTDL retains the gynoecium details of the flower, while most other methods blur the texture.

\textbf{Real MSI denoising}: In reality, an MSI may suffer from different kind of noises including not only the Gaussian noise but also the non-Gaussian noise, such as the stripe noise. Now we investigate the effectiveness of the proposed LTDL for recovering a real corrupted MSI. The denoising results are compared in Figure~\ref{fig:hsi}, where the 197th band of the HYDICE MSI has light noise and the 207th band has severe noise. The results of the LRTA, the PARAFAC and the Tdl are much poor than the others, which are not provided here owing to the space limitation. The BM4D is more appropriate for MSIs than the BM3D, and thus we also omit the results of BM3D to save the space. We zoom in a part of the restored image which involves stripe noise. It can be observed that although most methods achieve similar performance in the 197th band image that has light noise, the proposed LTDL is able to restore the 207th band which suffers from more complex and severe noise.

\par
\begin{figure}[t]
\begin{minipage}[b]{.52\linewidth}
  \centering
  \centerline{\includegraphics[width=4.5cm]{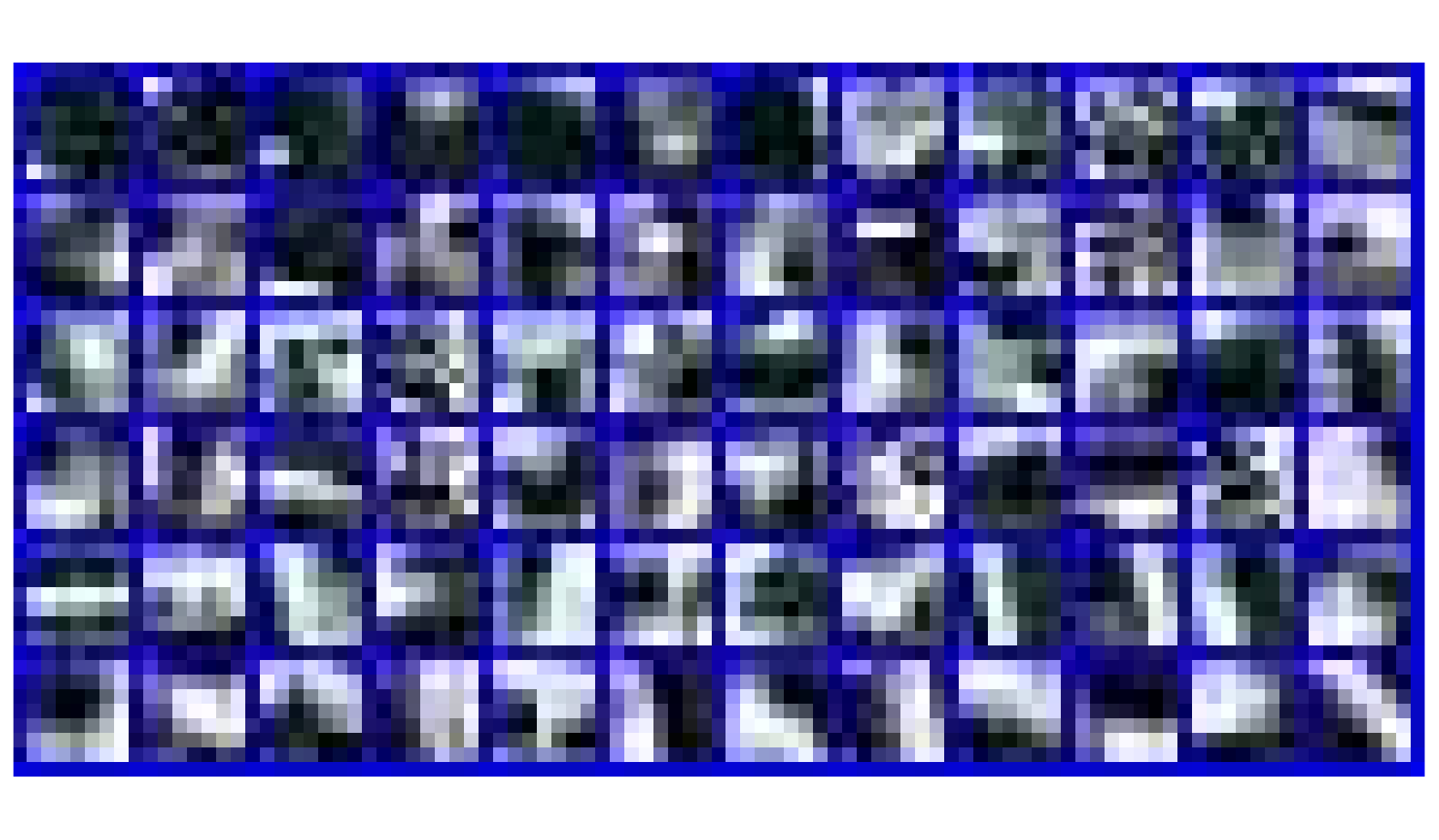}}
  \centerline{(a)}\medskip
\end{minipage}
\hfill
\begin{minipage}[b]{0.46\linewidth}
  \centering
  \centerline{\includegraphics[width=3.5cm]{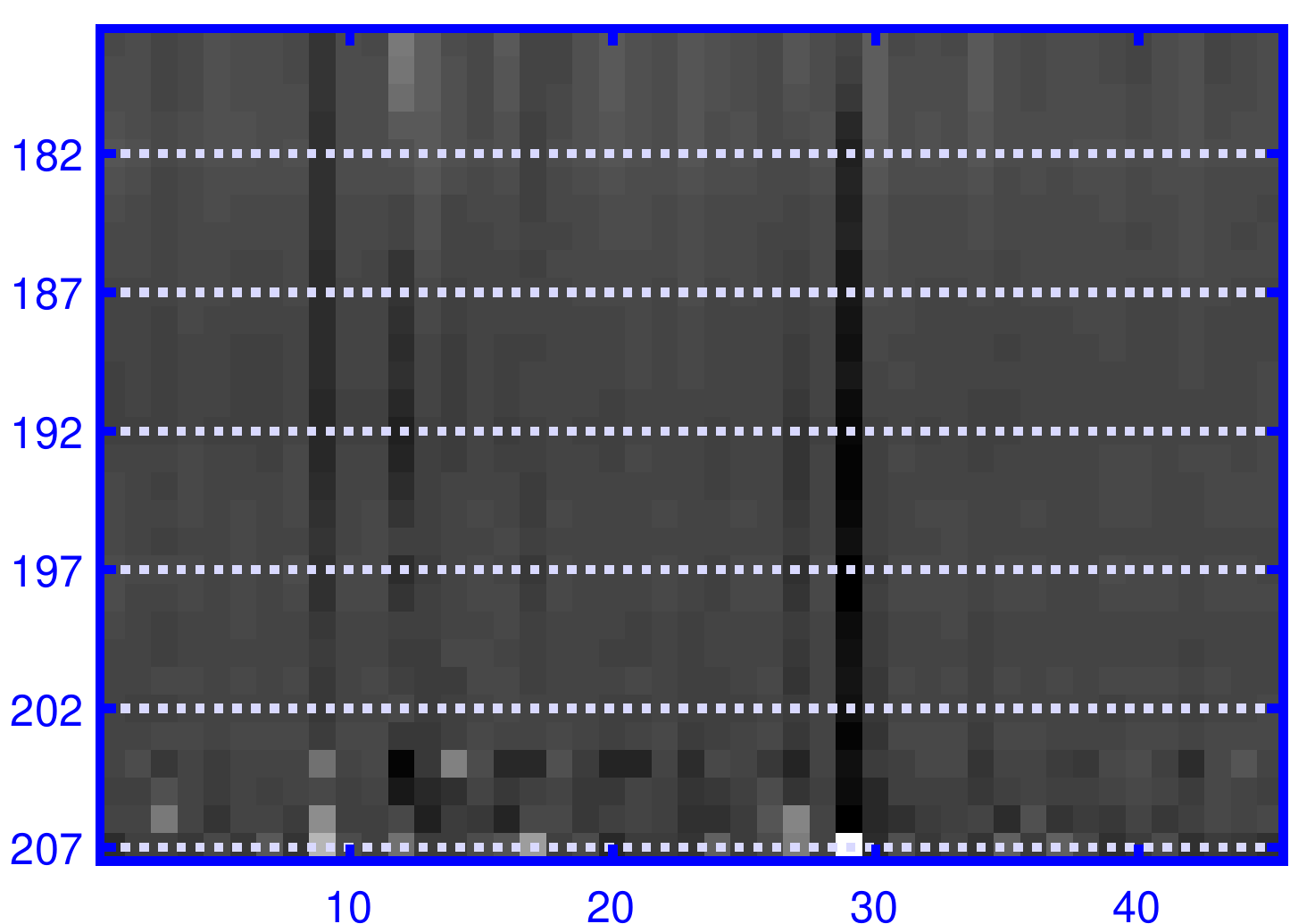}}
  \centerline{(b)}\medskip
\end{minipage}
\caption{Visualization of the learned dictionaries of the HYDICE MSI. (a) The learned spatial dictionary $D^a$; (b) The learned spectral dictionary $D^e$.}
\label{fig:dic}
\end{figure}

The learned dictionaries of the proposed method are shown in Fig. \ref{fig:dic}. Atoms in the spatial dictionary $D^a$ represent the spatial features of the HYDICE MSI. To enhance visualization, we reorganize each column into a patch of the size $7\times7$. Atoms of the learned spectral dictionary $D^e$ correspond to various spectral features of different bands.

%

\section{Conclusion}
\label{sec:Conclusion}
This paper presents an effective tensor dictionary learning method for restore high dimensional MSIs. The proposed LTDL method exploits the nearly low-rank structure in a group of similar blocks in the natural MSI and also exploits shared dictionaries among different groups, which makes the proposed method distinct to existing methods. Experimental results show the superior performance of the proposed method for denoising MSIs with both simulated corruptions and real corruptions.

{\small
\bibliographystyle{ieee}
\bibliography{egbib}
}

\end{document}